\documentclass[aps,twocolumn,amssymb,showpacs]{revtex4}

\usepackage{graphicx}

\newcommand{\be}{\begin{equation}}
\newcommand{\ee}{\end{equation}}
\newcommand{\bea}{\begin{eqnarray}}
\newcommand{\eea}{\end{eqnarray}}

\begin{document}

\title{The depletion in Bose Einstein condensates\\ using Quantum Field Theory in curved space}

\author{Roberto Balbinot$^a$, Serena Fagnocchi$^{b,a}$ and  Alessandro Fabbri$^c$ }

\altaffiliation{Email addresses: balbinot@bo.infn.it, fagnocchi@bo.infn.it,\\ afabbri@ific.uv.es}
\affiliation{{\it a)} Dipartimento di Fisica dell'Universit\`a di Bologna and INFN sezione di Bologna,  Via Irnerio 46,
40126 Bologna, Italy\\
{\it b)} Centro Studi e Ricerche {\it "Enrico Fermi"}, Compendio Viminale, 00184 Roma, Italy\\
{\it c)} Departamento de Fisica Teorica and IFIC, Universidad de
Valencia-CSIC,  C. Dr. Moliner 50, 46100 Burjassot (Valencia), Spain}



\begin{abstract}
Using methods developed in Quantum Field Theory in curved space we estimate the effects of the inhomogeneities and of a non vanishing velocity on the depletion of a Bose Einstein condensate within the hydrodynamical approximation.
\end{abstract}


\maketitle
\section{Introduction}
There is a 
growing interest in the so called \emph {analog models}, i.e. condensed matter systems
which, under some circumstances, behave as gravitational ones. The common feature of these
systems is that the fluctuations on top of the ground state propagate like fields moving on
a fictitious curved spacetime, whose metric tensor depends on the characteristics of the
underlying medium. This analogy holds both at the classical and at the quantum level. At the
moment this setting offers the most realistic  possibility to test peculiar effects of Quantum
Field Theory (QFT) in curved space which are still almost unobservable within the
gravitational context. Among these, the most popular one is surely Hawking's black hole
radiation \cite{hawking}. To this aim the   most promising candidates,  among analog models,
are the Bose Einstein condensates (BEC) \cite{stringari, BEC}, because
for them the actual experimental limits are very close to those necessary to observe such effects \cite{stime},
and the huge recent technological improvement in handling BEC makes us confident that this limit will be soon overtaken.\\
In this paper we use this analogy as a new tool in analytical calculations of the quantum depletion in dilute weakly interacting {\it non homogeneous} BEC. The depletion is the number density of the non condensed particles. In fact, because of quantum fluctuations, even at zero temperature a non vanishing fraction of particles are non condensed. This quantity has been analytically calculated  for homogeneous BEC only \cite{lee-huang} as function of the (constant) density of the condensate $n_0$. The most reliable attempt in extending the calculation  to non homogeneous BEC consists in considering the condensate locally as almost homogeneous, then replacing $n_0$ in the result obtained in \cite{lee-huang} by the actual non constant density $n$. This is the so called {\it local density approximation} (LDA) (see for ex. \cite{BEC.book}), which accounts for inhomogeneities only in first approximation, since no dependence on spatial derivatives of the density is considered. When the velocity is vanishing this approximation turns out to be quite good; however, as shown by the authors of \cite{somma-modi}, the LDA for a harmonic trap slightly overestimate the depletion.
Moreover, to our knowledge, nothing is known so far about effects due to the presence of a  non homogeneous velocity in the condensate.\\ In this paper we propose a framework, borrowed from QFT in curved space, which allows an explicit analytical evaluation of how inhomogeneities and velocity affect the depletion in a BEC.

\section{Quantum depletion in BEC}
In the usual s-wave scattering approximation, a 
Bose gas is described by a field operator $\hat \Psi$ satisfying
\be\label{eq.general}
i\hbar \frac{\partial}{\partial t}\hat \Psi=\left( -\frac{\hbar^2}{2m} \nabla^2+ V_{ext}(\vec x)+g \hat \Psi^\dagger \hat \Psi\right)\hat \Psi\, ,
\ee
where the coupling constant $g$ is related to the s-wave scattering length $a$ as $g=4\pi \hbar^2 a/m$, with $m$ the mass of the constituents of the gas and $V_{ext}$ the trapping external potential.
For a dilute gas, in the mean field limit, the field operator $\hat \Psi$ can be split into 
\be\label{U1}
\hat \Psi=\left(\Psi+\hat\varphi\right)\frac{\hat a}{\sqrt{\hat N}}\, ,
\ee
where $\Psi$ is the order parameter, $\hat\varphi$ the one particle fluctuation, $\hat a$ the distruction operator and $\hat  N=\hat a^\dagger \hat a$ the particles number operator. This splitting has been introduced following the so called {\it number conserving} approach \cite{castin,gardiner}. The splitting of the field $\hat \Psi$ as in Eq. (\ref{U1}) leads to $\langle \hat \Psi \rangle=0$: this preserves the U(1) symmetry of  Eq. (\ref{eq.general}). As a consequence, since $[\hat \varphi,\hat a/\sqrt{\hat N}]=0$, this framework keeps the total number of particles constant. 
However the use of a number conserving approach is not crucial for our purpouse, as will be discussed  below.\\
Neglecting the backreaction of fluctuations, the evolution of the condensate is governed by the Gross-Pitaevskii equation, which reads
 \be\label{GP}
i\hbar \frac{\partial}{\partial t} \Psi=\left( -\frac{\hbar^2}{2m} \nabla^2+ V_{ext}+g  n\right) \Psi\, ,
\ee
where $n=|\Psi^2|$. The fluctuation part satisfies the Bogoliubov-de Gennes equation
\be\label{B-deG}
i\hbar \frac{\partial}{\partial t}\hat \varphi=\left( -\frac{\hbar^2}{2m} \nabla^2+ V_{ext}+2 g n\right)\hat \varphi +g n \hat \varphi^\dagger
\, .\ee
The total particles density can be written as a sum of condensed and non condensed parts
\be
n_{tot}=\langle \hat \Psi^\dagger \hat \Psi\rangle=n+\langle \hat \varphi^\dagger \hat \varphi\rangle \, .
\ee
The last term describes the so called depletion.\\ It is worth noticing that the depletion depends on the Bogoliubov-de Gennes equation only. Therefore a framework which conserves the total particles number is not strictly necessary (al least at first order in $1/\sqrt{N}$, following the expansion of \cite{castin}). In fact the error induced by not using a number conserving approach is of the same order of that implicit in the Bogoliubov linearization of the theory. Instead of Eq. (\ref{U1}) one could split the field operator as $\hat \Psi=\Psi+\delta\hat\Psi$, with $\Psi=\langle\hat\Psi\rangle$  and $\langle\hat\Psi\rangle=0$, breaking the U(1) symmetry. This will lead to the same results \cite{carusotto}.\\
For a homogeneous condensate ($n=n_0=const$) at zero temperature this quantum depletion
can be exactly calculated \cite{lee-huang}:
\be\label{quantum.depletion}
\tilde n^q_0=\langle \hat \varphi^\dagger \hat \varphi\rangle =\frac{8}{3\pi^{1/2}} (n_0 a)^{3/2}\, .
\ee
At non zero temperature one has an additional contribution to the depletion because of thermal fluctuations. At low temperature ($\kappa_B T\ll n_0 g$, with $\kappa_B$ the Boltzmann constant), summing up both contributions, one can write the depletion for a homogeneous condensate as
\be\label{depletion.q+t}
\tilde n_0=\tilde n_0^q+\tilde n_0^T=\frac{8}{3\pi^{1/2}} (n_0 a)^{3/2}\left[1+\left(\frac{\pi \kappa_B T}{2 n_0 g}\right)^2 \right]\, .
\ee
It is worth noting that the derivation of the zero temperature quantum depletion (\ref{quantum.depletion}) is based on the full Bogoliubov dispersion relation for the modes, i.e. $\omega_k=c k \sqrt{1+\xi^2 k^2}$, where $c=(gn_0/m)^{1/2}$ is the velocity of sound and $\xi=\hbar/(\sqrt{2}mc)$ is the healing length of the condensate. Unlike this, the low temperature correction comes from
the phononic part of the spectrum only, i.e. $\omega_k=ck$. The integral giving the thermal depletion is dominated by modes with wavelength $\lambda \gg \xi$, so the result is insensitive to the high $k$ behavior of the dispersion relation. Therefore, while the correct description of the ground state requires the knowledge of the full "microscopic" theory \cite{foot1}
, deviations from the ground state value can, in some circumstances, be evaluated with sufficient accuracy using the low energy "macroscopic" phonon theory. This feature of the low energy theory is familiar in the Casimir effect. The standard Casimir energy density between two conducting plates can be evaluated with sufficient accuracy using the photon dispersion relation, provided the plate separation is much larger than the interatomic distance of the constituents of the plates \cite{kempf}.\\ 
Moving to non homogeneous condensates, when the 
current $\vec j\equiv \frac{1}{2im}\left( \Psi^*\vec \nabla \Psi-c.c \right)$ vanishes, the depletion is usually approximated by Eq. (\ref{depletion.q+t}) with now $n_0$ replaced by the actual space-time varying density $n(\vec x, t)$ in the  {\it local density approximation} (LDA) \cite{BEC.book}. We indicate this as
\be
\tilde n_{LDA}=\tilde n^q_{LDA}+\tilde n^T_{LDA}=\frac{8}{3\pi^{1/2}} (n a)^{3/2}\left[1+\left(\frac{\pi \kappa_B T}{2 n g}\right)^2 \right]\ ,
\ee
where the first term is the zero temperature quantum depletion and the second the thermal depletion, both in LDA. 

\section{BEC and the  gravitational analogy }

The aim of this paper is to show how, in view of the analogy described at the beginning, through the use of QFT in curved space and under some hypothesis, it is possible to obtain an estimate of the depletion which explicitly accounts for the inhomogeneities of the condensate (i.e. $\partial_i n\neq 0$) and for a non trivial velocity (i.e. a velocity that can not be simply eliminated by a Galileo transformation).\\ For BEC the analogy traces back to the hydrodynamical approximation of the basic equations for the condensate (\ref{GP}, \ref{B-deG}). In the density-phase representation for the condensate field $\Psi=\sqrt{n}e^{i\theta /\hbar}$, Eq. (\ref{GP}) is split in a continuity and an Euler equation, namely \cite{BEC.book}
\bea\label{idro.cont}
&&\frac{\partial}{\partial t}n+\vec\nabla\cdot (n\vec v)=0\, ,\\
\label{idro.euler}
&&m\frac{\partial}{\partial t}\vec v+\vec\nabla\cdot\left( \frac{mv^2}{2}+V_{ext}+gn-\frac{\hbar^2}{2m}\frac{\nabla^2\sqrt{n}}{\sqrt{n}}\right)=0\,
\ \ \ \eea
where $\vec v=\vec \nabla \theta/m$ is the velocity field. The last term in (\ref{idro.euler}) is called quantum pressure. The hydrodynamical approximation consists in neglecting this compared to the mean field term $gn$. For this approximation to be valid, the typical length scale $L$ for the variations of the condensate density $n$ has to be much bigger than the healing length ($L\gg\xi$). Decomposing the fluctuation operator $\hat\varphi$ as
\be
\label{acoustic.repr}
\hat \varphi=e^{i\theta/\hbar}\left( \frac{1}{2\sqrt{n}}\hat n_1+i\frac{\sqrt{n}}{\hbar}\hat \theta_1\right)\, ,
\ee
where $\hat n_1$ and $\hat \theta_1$ are real fields, one finds that for wavelengths
$\lambda\gg\xi$, the Bogoliubov-de Gennes Eq. (\ref{B-deG}) can be written as
\bea
\label{box}&\hat \Box
\hat \theta_1=0\,\, ,&\\
\label{n1}&\hat n_1=-g^{-1}\left(\partial_t\hat \theta_1+m^{-1}\vec \nabla \theta \cdot  \vec \nabla \hat \theta_1  \right)\, ,&
\eea where $\hat\Box=\frac{1}{\sqrt{-g}}\partial_\mu(\sqrt{-g}g^{\mu\nu}\partial_\nu)$
is the covariant d'Alembertian calculated from the so called "acoustic metric"
\be\label{gmunu} g_{\mu\nu}\equiv\frac{n}{mc}\left(
\begin{array}{cc}
-(c^2 -v^2)& -\vec v^T\\
-\vec v& \mathbf{1}
\end{array}\right)\ , \ee where $c$ is the local sound speed related to the density by $mc^2=gn$.
Hence the fluctuation $\hat \theta_1$ behaves {\it exactly} as a massless minimally coupled scalar field propagating on a fictitious curved spacetime described by the acoustic metric $g_{\mu\nu}$, whose line element reads
\be
ds^2=\frac{n}{mc}[-c^2 dt^2+(d\vec x-\vec v dt)\cdot(d\vec x-\vec v dt)]\, .\ee This is the core of the analogy \cite{review.visser}.\\For a homogeneous  condensate with constant velocity $\vec v_0$, the acoustic spacetime is flat as can be seen performing the Galileo transformation $d\vec x\rightarrow d\vec x-\vec v_0 dt$. This condensate configuration is a sort of ground state of the theory. On the other hand inhomogeneities or a non trivial velocity produce curvature. 
\\We will focus our analysis on the latter kind of configurations, looking for the corrections on the depletion induced by curvature with respect to the homogeneous (let say flat) result.
  
\section{QFT in curved space and the depletion in BEC}
As just mentioned,  phonons in a BEC propagate as a massless scalar field on a curved background spacetime: therefore they can  be described through the gravitational formalism. In the case of non homogeneous BEC, the curvature is not zero and we expect modifications with respect to the standard result induced by this curvature. \\In this case modes with $\lambda\stackrel{\sim}{<} \xi\ll L$  see the condensate as almost homogeneous. After renormalization, they will give the quantum depletion term in LDA, $\tilde n^q_{LDA}$. Deviations from this come from modes with wavelength $\lambda \stackrel{\sim}{>} L\gg\xi$. These are indeed the modes who feel the inhomogeneities of the condensate. They behave as a massles  scalar field on the curved spacetime $g_{\mu\nu}$. So the low energy phonon theory (even if unable to reproduce the ground state value of the depletion)  can   describe  with sufficient accuracy how inhomogeneities influence the depletion.  Now, within this spirit,  calling $\tilde n_{ph}$ the  depletion induced by the phonons, we can write $\tilde n =\tilde n^q_{LDA}+\tilde n_{ph}$, where, using Eq. (\ref{acoustic.repr}):
\bea\tilde n_{ph}&=&\langle \hat \varphi^\dagger\hat\varphi\rangle_{ph}=\frac{\langle
\hat n_1^2\rangle_{ph}}{4n}+\frac{n \langle \hat \theta_1^2\rangle_{ph}}{\hbar^2}\nonumber\\
&\approx& \frac{n \langle \hat
\theta_1^2\rangle_{ph}}{\hbar^2}\left(1+O(\frac{\xi^2}{L^2})\right)\, , \label{nin}\eea and the expectation
values are calculated using the phonon theory. The last approximation in Eq. (\ref{nin}) comes from taking $\hat n_1$ from Eq. (\ref{n1}) and estimating $\nabla\hat\theta_1\approx\frac{ \hat\theta_1}{L}$ and $\partial_t\hat\theta_1\approx\frac{ \hat\theta_1}{L/c}$.\\ In this way the calculation of
the depletion has been reduced to the calculation of the expectation value of the
squared of a massless scalar field minimally coupled to the acoustic spacetime
(\ref{gmunu}). $n\langle \hat \theta_1^2\rangle_{ph}/\hbar^2$ is dominated by wavelengths $\sim L$, therefore one can safely extrapolate the linear dispersion relation to high frequencies without affecting the finite renormalized result, since  the diverging terms appearing are then subtracted away.  \\ In QFT in curved space  a useful method of renormalization is point-splitting \cite{BD}. Within this procedure, $\langle \hat \theta_1^2(x)\rangle$ (here and in the following we will drop the subscript $ph$ from the expectation values) is the coincidence limit of the Euclidean Green function \be\langle \hat \theta_1^2(x)\rangle_{unren}=Re\left(\lim_{x'\rightarrow x}G_E(x,x')\right)\, . \ee Renormalization is performed by subtraction of the divergent part, the so called De Witt-Schwinger term $G_{DS}$, obtained  by the short-distance expansion of the Green function:
\be\langle \hat \theta_1^2(x)\rangle_{ren}=\lim_{x'\rightarrow x} (\langle \hat \theta_1^2(x,x')\rangle_{unren}-G_{DS}(x,x'))\, .\ee $G_{DS}(x,x')$ is purely geometrical, state independent and contains all the  short-distance divergences of the theory (for  technical details see for ex. \cite{BD}). 

\subsection{Static spherically symmetric configuration}

We now restrict our discussion to a static spherically symmetric configuration  ($n=n(r)$, $\vec v=0$).
The resulting acoustic metric is invariant under time-translation and $\partial_t$ is the Killing vector associated to this symmetry.
The acoustic metric (\ref{gmunu}) can be put in the standard form
\be\label{spher.symm.}
ds^2=-f(R)dt^2+h(R)dR^2+R^2d\Omega^2\, ,
\ee by the coordinate transformation  
\be\label{coord.transf.}
R=\sqrt{\frac{n}{mc}}\, \,r\, .\ee
Note that this coordinate transformation is valid only if $R\neq const$. The $R= const$ case will be treated separately.
 Now
\be \label{fh}f=\frac{nc}{m} \qquad h=(1+\frac{c'}{2c}r)^{-2}, \ee
and $'\equiv\partial_r$.\\ For configurations of this kind, it is
natural to consider thermal equilibrium states, i.e. whose Euclidean
Green function is periodic in $T^{-1}$ in the imaginary time. For
these states, Anderson et al. \cite{anderson} have developed a WKB
technique to evaluate $\langle \hat \theta_1^2(x,x')\rangle_{unren}$
in the case of a generic metric of type (\ref{spher.symm.}) leading
to \bea \label{phiquadro}
\langle\hat\theta_1^2\rangle_{ren}^{WKB}&=&\frac{\kappa_B^2
T^2}{12\hbar f}+\hbar\left[\frac{\mathcal{R}}{96\pi^2}\ln(\mu^2 f)
-\frac{f^{'2}}{96\pi^2 f^2 h}+\right. \nonumber\\&&\left.\right.
\nonumber\\&&\left.-\frac{f'h'}{192 \pi^2 f
h^2}+\frac{f''}{96\pi^2fh}+\frac{f'}{48\pi^2Rfh}\right] \eea
where only here $'\equiv\partial_R$, while $\mathcal{R}$ is  the curvature scalar and $\mu$  an arbitrary scale introduced by renormalization \cite{BD}.
The first term  describes a gas of massless scalar particles in thermal equilibrium at the temperature $T$ in a curved spacetime. The factor $1/f$ accounts for the "gravitational red-shift" (Tolman law \cite{frolovbook}). The other terms describe how the quantum  fluctuations are affected by the curvature of the effective spacetime.
Entering our expressions  (\ref{fh}) in Eq. (\ref{phiquadro}) one gets the following analytic expression for the thermal and inhomogeneity dependent part of the depletion:
\bea
\label{our.dep}
&&\tilde n_{ph}
=\nonumber\\
&&\frac{\kappa_B^2 T^2}{12\hbar^3 }\left(\frac{m^3}{ gn}\right)^{1/2}-\frac{(mgn)^{1/2}}{96\pi^2 \hbar }\left[ -\frac{7}{8}\frac{n'^2}{n^2}+\frac{5}{2}\frac{n''}{n}+\right.\nonumber\\ &&\left.+\frac{5}{r}\frac{n'}{n}\right]\ln 
\mu^2\frac{g^{1/2} n^{3/2}}{m^{3/2}}+\nonumber\\
&&+\frac{(mgn)^{1/2}}{48\pi^2 \hbar }\left[ -\frac{39}{16}\frac{n'^2}{n^2}+\frac{3}{4}\frac{n''}{n}+\frac{3}{2r}\frac{n'}{n}
\right]\, .
\eea
Note that the first  term in Eq. (\ref{our.dep}) elegantly reproduces  $\tilde n_{LDA}^T$. \\
The other terms give the corrections to the depletion in LDA induced explicitely by the spatial derivatives of the condensate density. They all vanish for a homogeneous BEC. 
\\One can obtain a numerical estimate of the relative weight of the inhomoheneities induced terms.  For $N=10^6$ Rb atoms trapped by a  harmonic  potential ($L_{osc}=1\mu m$, $a=5nm$, $R_0=L(15 a N/L)^{1/5}$ the size of the condensate) Eq. (\ref{our.dep}) gives at $T=0$ a correction with respect to the LDA result of $< 0.1\%$   at the centre of the trap, reaching $\sim 1\%$ approaching the boundary ($r\sim 0.9 R_0$). For $N=10^3 $ atoms the correction will be almost $2 \%$  at the center of the trap and almost $20\%$ at $r\sim 0.7 R_0$ \cite{mu}\cite{huang}. In both cases the corrections turn out to be negative, therefore lowering the actual number of non condensed particles. This is in qualitative agreement with what pointed out in \cite{somma-modi}.

\subsection{Radial velocity}
In the presence of a nonvanishing radial velocity (i.e. $\vec
v=v(r)\frac{\vec r}{|\vec r|}$) \cite{foot2} the system is no longer
in equilibrium and no analytical expression for
$\langle\hat\theta_1^2\rangle_{ren}$ is known. One can only evaluate how vacuum
fluctuations are modified by the presence of the radial velocity. Being the density and the velocity functions of the radial coordinate $r$ only,  one can  transform the acoustic metric (\ref{gmunu}) in the standard spherically symmetric form of Eq. (\ref{spher.symm.}) by the following coordinate transformation (assuming again $R\neq
const$) \be \tau=t+\int\frac{v\, dr}{c^2-v^2}\qquad\quad
R=\sqrt{\frac{n}{mc}}r\, ,\ee leading to a diagonal form of the type of Eq. (\ref{spher.symm.}) with now \be
\label{fh.v}f=\frac{n}{mc} (c^2-v^2)\qquad
h=\frac{c^2}{c^2-v^2}(1+\frac{c'}{2c}r)^{-2}. \ee The vacuum part of
Eq. (\ref{phiquadro}) for this new configuration gives \bea \label{our.dep.v} &&\tilde
n_{ph}^q=- \frac{m}{96\pi^2 \hbar c^2}\left[
\frac{3}{2}cc'^2+5c^2c''+-3v^2c''-2cv'^2-2cvv''\right.\nonumber\\
&&\left.+\frac{9}{2c}v^2 c'^2-4vc'v'+\frac{2}{r}(5c^2c'-4cvv'-v^2c')
\right.\nonumber\\ &&\left.-\frac{2}{r^2}v^2c\right]\ln \mu^2\frac{c(c^2-v^2)}{g}+\nonumber\\
&&+\frac{m}{48\pi^2 \hbar c^2}\left\{ -\frac{1}{2c(c^2-v^2)}(3c^2c'-v^2c'-2cvv')\right.\nonumber\\ &&\left.(3c^2c'-2v^2c'-cvv')-\frac{15}{4}cc'^2-\frac{1}{4c}v^2c'^2-\frac{5}{2}vc'v'+\right.\nonumber\\ &&\left.+\frac{3}{2}c^2c''-\frac{1}{2}v^2 c''-cv'^2-cvv''+\right.\nonumber\\
&&\left.+\frac{1}{r}(3c^2c'-v^2c'-2cvv')\right\}\, .
\eea
This equation generalizes $\tilde n_{ph}$ of Eq. (\ref{our.dep}) at $T=0$ in the presence of a radial velocity. Note that for homogeneous condensate ($c=const$) this expression does not vanish  for $|\vec v|=const$, as a consequence of the fact that a radial velocity can not be eliminated by a Galileo transformation. Eq. (\ref{our.dep.v}) reduces to (\ref{our.dep}) (at $T=0$) only for  $ v(r)=0$.\\
It is important to stress that Eq. (\ref{our.dep.v}) is strictly valid only in regions where the fluid velocity does not reach the sound velocity. For $|\vec v|=c$, Eq. (\ref{our.dep.v}) diverges.
This local divergence  has nothing to do with the ultraviolet divergences already canceled out in deriving Eq. (\ref{our.dep.v}). This divergence is quite familiar in  the gravitational context: 
in fact for $|\vec v|=c$ the $g_{00}$ term in the acoustic metric (\ref{gmunu}) vanishes and this is a hint of the fact that a horizon forms (the so called {\it sonic horizon}). This reflects on the modes of the scalar field that turn out to be singular  on the horizon \cite{our.rev}. As a direct consequence, a singular vacuum polarization on the horizon of a black hole appears. However it is well known that taking  the black hole formation process into account a Hawking radiation emerges: its presence is responsible for an additional contribution to $\langle\hat\theta_1^2\rangle_{unren}$ which has an equal and opposite sign divergence, leading to a $\langle\hat\theta_1^2\rangle_{ren}$ regular on the horizon. 
This has been explicitely shown for a Schwarzschild black hole (for which $f(r)=g^{-1}(r)=1-2M/r$) \cite{frolovbook}.
A similar behavior is expected also in the BEC case.
\\ Unfortunately for a general metric as in Eq. (\ref{spher.symm.}) an explicit analytic form for the $\langle\hat\theta_1^2\rangle_{unren}$ taking  Hawking radiation into account and valid in the near horizon region is still unknown.

\subsection{$R$=constat case}

For the sake of completeness let us consider the case when the coordinate transformation (\ref{coord.transf.}) is singular, i.e. when $n/c\propto 1/r^2$.  The acoustic metric (\ref{spher.symm.}) shows that the surface $(r,t)=const$ has constant area. In this case Eq. (\ref{phiquadro}) is replaced by
 \bea
\label{phiquadro.Rconst}
\langle\hat\theta_1^2\rangle_{ren}^{WKB}&=&\frac{\kappa_B^2
T^2}{12\hbar f}+\hbar\left[\frac{\mathcal{R}}{96\pi^2}\ln(\mu^2f)
-\frac{f^{'2}}{96\pi^2 f^2 h}+\right. \nonumber\\&&\left.\right.
\nonumber\\&&\left.-\frac{f'h'}{192 \pi^2 f
h^2}+\frac{f''}{96\pi^2fh}\right] \eea 
with $'=\partial_r$, $
h=\frac{nc}{m} (c^2-v^2)^{-1}$ and $f$ unchanged as in Eq. (\ref{fh.v}). \\For the standard equation of state for BEC $mc^2=gn$, one can take $n=n_0 r_0^4/r^4 $, with $n_0$ and $r_0$ two constants having the dimension of density and length respectively.
The
resulting  quantum depletion reads
\bea 
n_{ph}^q&=&\frac{r_0^2 }{48\pi^2 \hbar r^4}(n_0 g m)^{1/2}\left\{-\left(10+\frac{v}{c}-\frac{r^2}{c^2}(vv')'\right)\right.\nonumber \\
&&\times\ln
\left(\mu^2\frac{g^{1/2}n_0^{3/2}}{m^{3/2}}\frac{r_0^6}{r^6}\right)-\nonumber \\
&&-(1-\frac{v}{c})^{-1}\left( 21-22\frac{v^2}{c^2}+5\frac{v^4}{c^4}+\right.\nonumber\\ 
&&\left. r \frac{vv'}{c^2}(10-8\frac{v^2}{c^2})+r^2 \frac{v^2v'2}{c^4}\right)\nonumber\\
\label{our.depq.Rconst}
&&\left.+21-3\frac{v^2}{c^2}+4r\frac{vv'}{c^2}-r^2\frac{(vv')'}{c^2}\right\}\, .
\eea

\section{Conclusions}
Summarizing, we have shown how sophisticated renormalization techniques originally developed to study
quantum effects in black hole physics, can be extremely useful in a (seemingly)
complete different context, namely BEC.
In this paper these techniques have been used to give an analytical estimation of the effects of spatial derivatives and a non trivial velocity on the depletion of a BEC. Our predictions are  beyond the present experimental capability. Nonetheless this work represents the first attempt to go beyond the LDA, providing a general analytical scheme that can be used in cold-atoms systems within the hydrodynamical approximation.

{\bf Acknowledgments:} the authors would like to thank I. Carusotto, P. Pieri, A. Recati and S. Stringari for stimulating discussions and for their  useful comments on the manuscript. S.F. also thanks "E. Fermi" Center for supporting her research.


\begin{thebibliography}{99}
\bibitem{foot1}The linear phonon dispersion relation, characteristic of a massless relativistic free field would lead (after renormalization) to a vanishing quantum depletion.
\bibitem{foot2} To realize this kind of configurations, it is necessary to add source (well) terms in the continuity Eq. (\ref{idro.cont}) in order to conserve the mass. However the presence of these terms do not invalidate our treatment since Eqs. (\ref{box}, \ref{n1}) remain unchanged.
\end{thebibliography}
\end{document}